\documentclass[aps,prl,reprint,superscriptaddress,nofootinbib,floatfix,showkeys]{revtex4-2}
\usepackage[top=1.0in, bottom=0.75in, left=0.7in, right=0.7in]{geometry}
\usepackage[utf8]{inputenc}
\usepackage[T1]{fontenc}
\usepackage{amsmath}
\usepackage{amssymb}
\usepackage{ulem}
\allowdisplaybreaks
\usepackage{graphicx}
\usepackage{multirow}
\usepackage{dcolumn}
\usepackage{xcolor}
\usepackage{natbib}
\usepackage{float}
\usepackage{bm}
\usepackage{comment}
\usepackage{tabularx}
\usepackage{booktabs}
\usepackage[italicdiff]{physics}
\usepackage[colorlinks=True,allcolors=blue]{hyperref}
\usepackage{orcidlink}
\usepackage{soul}
\usepackage{float}
\usepackage{geometry}

\definecolor{darkgreen}{rgb}{0, 0.5, 0.3}
\usepackage{siunitx}
\newcommand\redsout{\bgroup\markoverwith{\textcolor{red}{\rule[0.5ex]{2pt}{0.4pt}}}\ULon}
\newcommand\IITRxPH{Department of Physics,
Indian Institute of Technology Roorkee, Roorkee 247667, India}

\newcommand\IITRxCPQCT{Centre for Photonics and Quantum Communication Technology,
Indian Institute of Technology Roorkee, Roorkee 247667, India}

\begin{document}

\title{Scalable nuclear shell model calculations on noisy quantum computers}

\author{Durgesh Pandey\,\orcidlink{0009-0001-0877-1980}}
\email{durgesh_p@ph.iitr.ac.in}
\affiliation{\IITRxPH}
\author{Ankit Kumar Das\,\orcidlink{0009-0001-2252-2657}}
\email{ankitk_das@ph.iitr.ac.in}
\affiliation{\IITRxPH}

\author{P. Arumugam\,\orcidlink{0000-0001-9624-8024}}
\email{arumugam@ph.iitr.ac.in}
\affiliation{\IITRxPH}
\affiliation{\IITRxCPQCT}

\begin{abstract}
The exact diagonalization of the nuclear shell model scales exponentially, leading to severe memory bottlenecks in classical high-performance computing (HPC). While hybrid quantum algorithms like the Variational Quantum Eigensolver (VQE) aim to overcome these limits, their deep quantum circuits and iterative feedback loops are susceptible to substantial noise inherent in the current Noisy Intermediate-Scale Quantum (NISQ) hardware. This noise renders several algorithms, such as the VQE, impractical for large-scale calculations despite sophisticated noise-mitigation techniques. As a pragmatic approach tolerant to these issues, we apply the Sample-based Quantum Diagonalization (SQD) framework to nuclear shell models for the first time. Using $^{38}\text{Ar}$ as a benchmark to confirm the numerical accuracy, we extend SQD to $^{32}\text{Mg}$, solving a nuclear shell-model Hamiltonian whose underlying Hilbert space cannot be directly diagonalized using conventional classical methods in a given HPC system. We present a systematic comparison of SQD with standard variational quantum schemes and exact classical solvers. By leveraging NISQ hardware connected via the cloud to classical HPC clusters, the SQD-based scheme could outperform conventional supercomputers in memory scaling and total execution time, enabling more rigorous large-scale shell model calculations.
\end{abstract}

\maketitle

\textit{Introduction}---It is possible to overcome some resource limitations of classical computers by offloading certain tasks to quantum computers, thereby achieving quantum advantage \cite{AJAGEKAR2020106630,hep_quantum_advantage}.  This approach is useful for simulating many-body quantum systems, a severe computational challenge that demands immense processing power and storage capacity~\cite{troyer2005computational}. The Hilbert-space dimension exhibits exponential growth with the number of spin-orbitals ($N$), scaling as $\mathcal{O}(2^{N})$.  Consequently, memory requirements can increase dramatically; in many exact-diagonalization approaches, adding one additional degree of freedom approximately doubles the required storage~\cite{noack2005diagonalization, wietek2025xdiag}. Due to this drastic escalation in resource needs, even high-performance computers struggle to determine the necessary ground and excited states when simulating these systems by solving the Schrödinger equation for a large Hamiltonian matrix. While various numerical techniques have been engineered to solve sparse matrices, including specialized Hermitian and Lanczos solvers for eigenvalue computation~\cite{lanczos1950iteration, nyman2001iterative}, each approach faces its own limitations. Ultimately, the defining bottleneck across all these methods remains the exponential growth of the Hilbert space dimension, leading to severe memory and computational scaling challenges~\cite{nandy2025quantum, schollwock2005density}.

Some problems can be solved with a truncated Hilbert basis~\cite{Modine1996truncation}, but this approach is not applicable to systems where interactions between different levels are highly important, leading many-body systems to exist as superpositions of many configurations. The atomic nucleus is such a highly correlated system rather than a collection of independent particles~\cite{hergert2020abinitio}. As the number of nucleons increases, particularly where traditional magic shell closures break down far from stability, these nuclear systems show even stronger many-body correlations and collective behavior~\cite{caurier2005shellmodel}. In these cases, the simple closed-shell (inert-core) approximation is inadequate. Because of significant configuration mixing and the growing involvement of core excitations, we need to use a much larger configuration space. A benchmark example of this breakdown of the core-valence separation in neutron-rich nuclei such as $^{32}\text{Mg}$~\cite{otsuka2005tensorforce}.  Although it has $N = 20$, a traditional magic number in the nuclear shell model, it does not behave like a closed-shell system~\cite{BARRETT2013131}. Instead, it lies within the ``Island of Inversion,''~\cite{utsuno1999varying, shimizu2019thick}, where the expected shell gap shrinks, and intruder configurations from higher orbitals become energetically favored. As a result, neutrons are pushed across the shell gap, leading to strong configuration mixing and deformation. This produces clear collective features, such as rotational spectra and enhanced quadrupole moments. To properly study $^{32}\text{Mg}$, one must go beyond the simple core-plus-valence approach and use large-scale shell-model calculations that include multi-particle-multi-hole excitations. The core can no longer be treated as inert, as it actively participates through polarization and excitations. This makes $^{32}\text{Mg}$ a strict test for effective interactions and modern many-body methods. Its behavior highlights the importance of correlations, tensor forces, and the changes in shell structures far from stability. Consequently, $^{32}\text{Mg}$ plays a key role in refining our understanding of nuclear forces and the limits of shell closures~\cite{kitamura2021coexisting, Warburton1990_32region_29to44}.

For $^{32}\text{Mg}$, an exact calculation using a no-core shell model requires a Hilbert space with a dimension of $D \gtrsim 2^{64}$, which is already far too large to calculate. Even assuming a frozen, inert $^{16}\text{O}$ core, including the full $sd$-$pf$ cross-shell excitations, leaves us with an extremely large basis dimension of around $D \sim 2^{48}$.  This makes exact diagonalization impossible because any calculation that scales as $D^2$ cannot be performed on a computer~\cite{georgescu2014quantum, fauseweh2024quantum}. 
 
Therefore, whether we use a no-core framework or a full valence space $sd$-$pf$~\cite{utsuno1999varying}, the exact solutions are completely out of reach.  To mitigate this issue, we propose a quantum-classical hybrid framework based on Sample-based Quantum Diagonalization (SQD)~\cite{Barison_2025_SQD, Duriez2026}, which leverages real quantum hardware more efficiently than other algorithms. 

Recently, exact shell-model calculations for $^{32}$Mg have been proposed using Quantum Phase Estimation (QPE)~\cite{benstead2026QPE32Mg}, a framework for fault-tolerant devices that is therefore not feasible on current Noisy Intermediate-Scale Quantum (NISQ) hardware. Similarly, Quantum Krylov-based methods have been investigated for shell-model calculations of other nuclei~\cite{yoshida2026nuclearmanybodysystemsQPEQKrylov}. In parallel, variational approaches, including the Variational Quantum Eigensolver (VQE), its variations ~\cite{Singh_Nifeeya_shellmodel_GC}, Quantum Generator Coordinate Method (QuGCM), and related shell-model algorithms, have also been developed for nuclear structure calculations~\cite{pandey2026quGCM}.  The practical implementation of these methods often depends on future advances in quantum hardware, whereas the SQD framework employed in this work operates efficiently on current NISQ hardware, thereby bringing practical quantum advantage closer to reality for large-scale shell-model calculations.

\begin{figure*}[th]
    \centering 
    \includegraphics[width=1\linewidth]{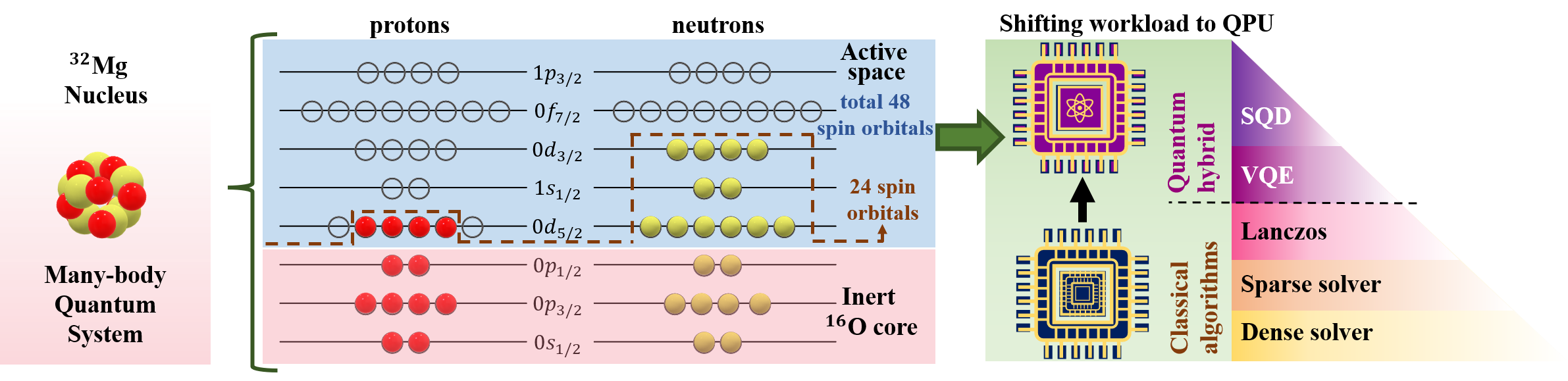}
        \caption{The schematic diagram of a nuclear many-body system and the basic idea to shift workloads to a QPU with increasing configurations. The triangular hierarchy at right depicts the increasing degree of quantum participation in shell-model calculations. Dense, sparse, and Lanczos methods correspond to classical computation, while VQE and the proposed SQD progressively shift more of the workload to the QPU. The SQD framework achieves the best overall computational performance for the largest problems investigated, demonstrating the utility of NISQ hardware.}
    \label{fig:Graphical abstract}
\end{figure*}


\textit{Formalism}---To suit simulations in quantum computers, the Hamiltonian constructed through single-particle energies and two-body matrix elements describing the interaction is expressed in terms of creation and annihilation operators, popularly known as the second quantization form~\cite{fetter_2003_Quantum_theory_of_many_particle_systems}, as 
\begin{equation}
\label{H_2nd_quant}
    \hat{H}
    =
    \sum^{N-1}_{p,q=0} h_{pq}\, a_p^\dagger a_q
    +
    \frac{1}{2}
    \sum^{N-1}_{p,q,r,s=0}
    h_{pqrs}\,
    a_p^\dagger a_q^\dagger a_r a_s .
\end{equation}
This second-quantization form is then converted to the `SparsePauli' form using the One-Hot (OH) encoding corresponding to the Jordan-Wigner (JW) transformation, which can be directly aligned over a quantum circuit that can run on a quantum computer~\cite{JordanWigner1928, Whitfield10032011, Batista2001}: 
\begin{equation}
    \hat{H}
=
\sum_{i} h_i P_i ,
\end{equation}
where \(h_i \in \mathbb{R}\) are the real coefficients corresponding to the Pauli matrices $P_i \in \{I, X, Y, Z\}^{\otimes N}$ denoting the $N$-qubit Pauli operators.

To construct the ansatz, or trial function, we begin with a reference state. The reference state $\ket{\phi_\text{HF}}$ is the minimum energy Hartree-Fock state, represented as an $N$-bit occupation string, where a value of $1$ denotes an occupied spin orbital and $0$ denotes an unoccupied spin orbital. Using the Jordan-Wigner (JW) transformation, the fermionic occupation-number basis is mapped directly onto the computational basis of an $N$-qubit register~\cite{JordanWigner1928, Batista2001}. Each spin orbital is represented by a single qubit, where the states $\ket{0}$ and $\ket{1}$ denote unoccupied and occupied orbitals, respectively. Hence, a Slater determinant such as $\ket{10001000\cdots}$ is initialized from $\ket{0}^{\otimes N}$ by applying Pauli-$X$ gates to the qubits corresponding to the occupied spin orbitals~\cite{Whitfield10032011,Batista2001}. Consequently, the quantum state resides in the $2^N$-dimensional Hilbert space of the $N$-qubit register.

The variational state is prepared by applying the Unitary Coupled Clusters Singles and Doubles (UCCSD)~\cite{Bartlett_Rodney2006_New_perspectives_on_unitary_coupled_cluster_theory,Romero_2019_UCCansatz} operator to this Hartree-Fock reference state:
\begin{equation}
\label{Eq:Psi_UCCSD}
|\Psi\rangle = G_\text{UCCSD}(\boldsymbol{\theta}) |\phi_\text{HF}\rangle ,
\end{equation}
where the unitary operator $G_\text{UCCSD}$ is defined as
\begin{equation}
\label{UCCSD_op}
G_\text{UCCSD}(\boldsymbol{\theta})
=
e^{\,T(\boldsymbol{\theta}) - T(\boldsymbol{\theta})^\dagger}.
\end{equation}
Here, the cluster operator $T(\boldsymbol{\theta})$ contains the single and double excitation operators that move particles from the occupied $N_f$ orbitals to the vacant ones to incorporate configuration mixing and core excitations. It is explicitly given by:
\begin{equation}
T(\boldsymbol{\theta})
=
\frac{1}{2}
\sum^{N-1}_{a=N_f}
\sum^{N_f-1}_{j=0}
\theta_{j}^{a}\,
a_a^\dagger  a_j 
+
\frac{1}{4}
\sum^{N-1}_{a,b=N_f}
\sum^{N_f-1}_{j,k=0}
\theta_{jk}^{ab}\,
a_a^\dagger a_b^\dagger a_k a_j .
\end{equation}

Being unitary, $G_\text{UCCSD}$ is compiled into a sequence of quantum gates. For particular optimized values of the parameters $\boldsymbol{\theta} \equiv \{\theta_j^a,\theta_{j,k}^{a,b} \}$ that include relevant terms from the Hamiltonian [Eq.~(\ref{H_2nd_quant})], $\ket{\Psi}$ converges to the correct eigenvalues representing the energy levels for the many-body quantum system under consideration. 

In VQE \cite{Dumitrescu2018Cloud_Quantum_Computing_of_an_Atomic_Nucleus, peruzzo2014variational, Perez-Obiol2023Nuclear_shell-model_simulation_in_digital_quantum_computers}, to get the ground state energy, the parameters are obtained through a classical optimizer based on the measurement of $\bra{\Psi}\hat{H}\ket{\Psi}$ by executing a quantum circuit made from the ansatz and the `SparsePauli' form of $\hat{H}$. This method has several caveats, including noise and continuous HPC-QPU feedback, that could be avoided in the approach given below.

\begin{figure*}
    \centering
    \includegraphics[width=0.9\linewidth]{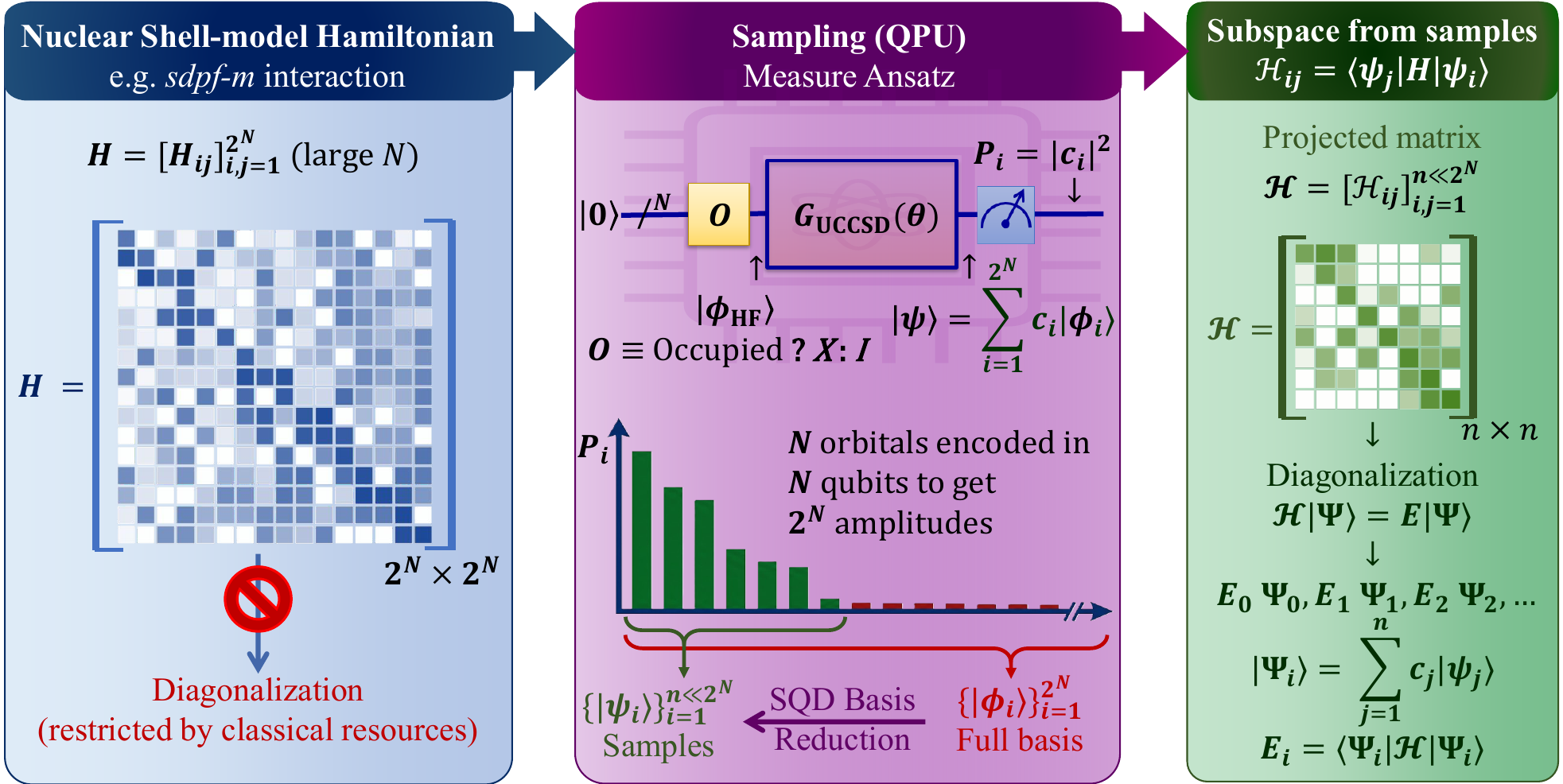}
    \caption{Illustration of a simplified SQD workflow: constructing the ansatz from the nuclear Hamiltonian, sampling from a quantum circuit on real quantum hardware, filtering the results (with noise removal, bitstring flipping, etc., to recover particle-number conserved configurations), and ultimately creating a subspace from the recovered samples. The QPU can perform the necessary sampling to identify states that contribute to the ground state, whereas classical computers struggle as resources grow exponentially. The full post-processing step of SQD for configuration recovery is omitted for simplicity.}
\end{figure*}

\textit{Sample-based Quantum Diagonalization (SQD)}---This method was developed to compute ground-state energies at scales beyond classical capabilities by treating the quantum hardware solely as a configuration sampler~\cite{Barison_2025_SQD, patra2026towards, SQD_2025_Chemistry_beyond_the_scale_of_exact_diagonalization_on_a_quantum-centric_supercomputer}. This foundational framework has subsequently been expanded to include sample-based Krylov diagonalization techniques~\cite{Huggins_2020_IOP_nonorthogonal_VQE_solver}. The process begins with the execution of a shallow, hardware-efficient quantum ansatz to approximate the target wave function. Unlike variational algorithms, this method does not require precise optimization of the quantum state. Instead, the prepared state only needs to overlap sufficiently with the low-energy physical subspace. A non-negligible overlap is enough for the algorithm to work effectively. Therefore, for sampling, we consider a random superposition of different states from a few UCCSD excitations over the Hartree-Fock state. These excitations are governed by the Hamiltonian to include a few of the possible excitations. This naturally extends to give other possible excitations during the measurement of this state. The quantum processing unit (QPU) measures this prepared state across a large number of shots, producing a probabilistic dataset of measurement bitstrings corresponding to specific fermionic configurations. With the inputs from QPU, the classical system distills the astronomically large full Hilbert space into a concentrated active subspace. This raw dataset is then transferred to a classical computing environment, where a physically motivated truncation is applied by discarding configurations that are affected by noise (values that fall below a specific measurement-frequency threshold). The physically impossible results are also excluded by particle-number conservation and symmetry.  Following the construction of this active subspace, the target Hamiltonian $\hat{H}$ is classically projected onto the truncated basis. The classical HPC system performs exact diagonalization on the resulting reduced matrix to simultaneously extract the ground state and low-lying excited states. To recover any critical physical configurations discarded during the initial truncation or lost to quantum noise, SQD utilizes a classical self-consistent iteration. Using the average orbital occupancies derived from the diagonalized state, new configurations are probabilistically generated and appended to the active space. The matrix projection and diagonalization steps are then iterated until rigorous energy convergence is achieved. 

By decoupling state optimization from the quantum hardware, SQD fundamentally eliminates the latency-heavy feedback loops and barren plateau vulnerabilities characteristic of variational algorithms~\cite{Wang2021_barren_plateaus} like the VQE and Variational Quantum Deflation (VQD)~\cite{Higgott2019variationalquantumdeflation, Singh_Nifeeya_shellmodel_GC}. Furthermore, SQD has intrinsic resilience to hardware noise without requiring resource-intensive Quantum Error Mitigation (QEM) protocols. Quantum hardware noise primarily affects the selection of basis states, while the eigenvalues are ultimately obtained via classical diagonalization in the low-dimensional subspace spanned by the selected basis. This selection naturally tends to favor the dominant basis states, and hence, hardware noise leads to the discarding of some weakly contributing states.  Consequently, the results for important low-energy eigenstates remain accurate.

Compared with existing quantum hybrid approaches, the SQD selects which basis to keep in the ground-state configuration, effectively truncating the basis. In contrast, VQE only balances the superposition of the states under consideration to obtain the exact ground state.
One should be cautious that SQD cannot be efficient if full interactions are needed; it works well if the answer lies within the subspace of the full configuration. The ground state approximated through SQD is given by
\begin{equation}
    \ket{\Psi_{\mathrm{gs}}}
    =
    \sum_{j=1}^{n \ll 2^N} c_j \ket{\psi_j}
    \approx
    \sum_{i=1}^{2^N} c_i \ket{\phi_i},
\end{equation}
where $\{\psi_j\}\subset \{\phi_i\}$.
VQE works for full configurations as well as truncated ones, where ansatzes are constructed to include the required basis to improve optimization. It determines the constants $c_i$, sometimes called weights, by minimizing the energy on a quantum computer with respect to them.  SQD can also find lower excited states, as the final diagonalization is done classically, whereas VQE and related methods can only find the ground state and require updates to determine other energy levels such as VQD, Variational-VQE (VVQE) and modifications for non-hermitian open-systems~\cite{pandey2026rvvqe}, or Subspace-search VQE (SSVQE)~\cite{Danbo20222VVQE, Nakanishi2019_SSVQE}. These methods, however, require re-running Quantum circuits or increasing their depth, making the solution less accurate and more resource-consuming or time-consuming.


\begin{figure*}[t!]
   \centering    \includegraphics[width=0.9\linewidth]{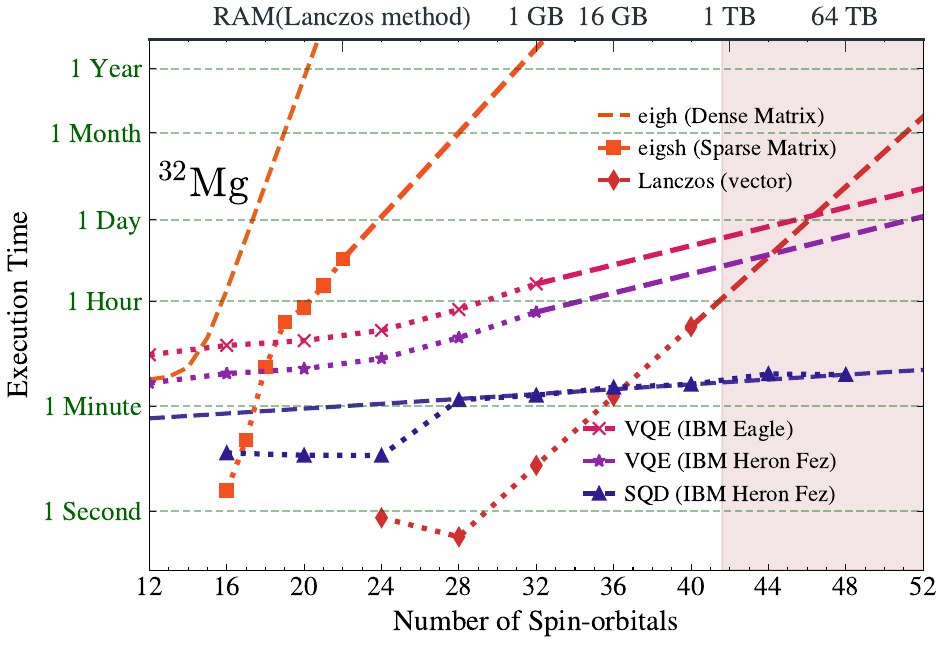}
    \caption{Comparison of the computational scaling of various classical algorithms and the hybrid quantum algorithms VQE and SQD for the shell-model calculations of $^{32}$Mg. The dotted lines connect the computed results (shown by various symbols), and the dashed lines are theoretical extrapolations for each method. The top axis indicates the approximate memory (RAM) required by the Lanczos method as the number of spin-orbitals increases. The region beyond the RAM limit ($768$~GB) of a high-memory compute node in the PARAM Ganga ~\cite{paramganga} high-performance computing system used in this work is shaded in red. For a given number of spin-orbitals ($N$), the full Hilbert space size is $2^N\approx 10^{0.3N}$. }
    \label{fig:quantumadvflowchart}
\end{figure*}

\textit{Results}---Recent literature has demonstrated the efficiency of variational methods for determining energy levels in nuclear systems, like VQE for a few energy levels corresponding to the $^{38}$Ar nuclei, which is in good agreement with the classical results~\cite{Singh_Nifeeya_shellmodel_GC}. Using this well-known and proven method in quantum computing as a baseline, we can determine and compare the accuracy and time advantages of other methods for finding the ground-state energy.  

We modeled the \(^{38}\)Ar nucleus as a \(^{28}\)Si core with two holes in the \(sd\)-shell. The valence space consists of the \(s_{1/2}\) and \(d_{3/2}\) orbitals. We used the USDB interaction to construct the Hamiltonian for diagonalization, allowing us to determine the ground and excited states~\cite{RichterUSDA_USDB,Singh_Nifeeya_shellmodel_GC}.
In Table~\ref{tab:Ar38SQDresults} we present our SQD results for $^{38}$Ar, which perfectly reproduce the classical exact diagonalization values. Crucially, SQD demonstrates a massive enhancement in computational efficiency, recovering both the ground and the first excited states in just $3.16$ seconds. Details of the computational time estimate are provided in the literature~\cite{PoojaRao_model_timecalc2026}. In contrast, VQE required approximately $199.53$ seconds for the ground state and an additional $203.57$ seconds for the first excited state. In VQE, the expectation value of the Hamiltonian must be evaluated iteratively on the QPU during optimization. For simulation, the QPU time for VQE is estimated for longer runs~\cite{PoojaRao_model_timecalc2026}, while it is confirmed for shorter circuits with actual runs. Since optimization typically requires hundreds of iterations, quantum hardware repeatedly executes deep quantum circuits, leading to a longer time and the accumulation of statistical errors and hardware noise in the estimated eigenvalues. In contrast, SQD's high efficiency in using the QPU enables real QPU execution time in all spin-orbital cases. The SQD accesses the QPU only once for sampling, with a sufficient number of shots (Details are provided in the supplementary material). Having validated SQD on $^{38}$Ar, we scale the methodology to $^{32}$Mg, again demanding rigorous agreement with classical methods. 

\begin{table}[htb]
\centering
\caption{Comparison of the energies of the ground ($0_1^+$) and first excited ($0_2^+$) states of $^{38}$Ar obtained from exact diagonalization (Classical), VQE~\cite{Singh_Nifeeya_shellmodel_GC}, and SQD under noiseless and noisy conditions. Numbers in parentheses denote the uncertainty in the last digits of the corresponding energies arising from repeated runs.}
\label{tab:Ar38SQDresults}
\begin{tabular}{lcc}
\toprule
\textbf{Method} & $\mathbf{0_1^+}$ & $\mathbf{0_2^+}$ \\
\midrule
Classical        & $-152.677$      & $-149.225$ \\
VQE (Noiseless)\footnote[1]{With an ideal quantum simulator without hardware noise.}  & $-151.794(448)$ & $-137.468(187)$ \\
VQE (Noisy)\footnote[2]{With real quantum hardware where the uncertainties include finite-shot sampling and hardware noise.}      & $-139.873(942)$ & $-110.873(306)$ \\
SQD (Noiseless)\footnotemark[1]  & $-152.677(206)$      & $-149.225(319)$ \\
SQD (Noisy)\footnotemark[2]      & $-152.144(381)$      & $-149.199(827)$ \\
\bottomrule
\end{tabular}
\end{table}

Assuming $^{16}$O as an inert core for $^{32}$Mg and considering configuration interactions sdpf-m~\cite{utsuno1999varying}, we calculate the ground-state energy with an increasing number of spin orbitals in our configuration-interaction basis using the PARAM Ganga HPC system at IIT Roorkee \cite{paramganga}, within its Python environment. A single high-memory compute node is used for fair evaluation and comparison. Classical calculations begin with dense diagonalization (eigh)~\cite{2020SciPy}, where the complete Hamiltonian matrix is explicitly stored. The sparse eigensolver eigsh~\cite{2020SciPy} extends the accessible problem size by storing only nonzero matrix elements. For even larger model spaces, a matrix-free Lanczos implementation~\cite{Lanczos__method,lanczos1950iteration} is adopted, avoiding explicit storage of the Hamiltonian by evaluating Hamiltonian-vector products on demand through a Numba-accelerated kernel. This provides the largest classically tractable calculations before transitioning to the quantum-hybrid VQE and SQD frameworks.

\begin{table*}[htb!]
\caption{Evolution of the effective valence ground-state energy (in MeV) of $^{32}$Mg calculated as a function of the total active spin-orbitals using a frozen $^{16}$O core framework. Systematically increasing the model-space size improves agreement with the experimental value $-122.012$~MeV \cite{Huang_2021, Wang_2021_II} and $-117.58$~MeV from the ab initio study in the island of inversion~\cite{Cao_2025}. The label `OOM' (Out of Memory) denotes configurations in which the dimension of the ground-state Hamiltonian matrix required for classical diagonalization exceeded the available $768$~GB RAM threshold.}
\label{tab:energy_evolution}
\centering
\begin{tabular}{cccr@{(}lcS[table-format=-3.4(4)]c}
\toprule
\toprule
\textbf{Spin}
& \multicolumn{4}{c}{\textbf{Dimension of Hilbert Space}}
& \textbf{Shots}
& \multicolumn{2}{c}{\textbf{Ground-state Energy (MeV)}} \\

\cmidrule(lr){2-5}\cmidrule(lr){7-8}

\textbf{orbitals $(N)$}
& \quad\textbf{Full} $(2^N)$
& \quad\textbf{Particle-number conserved}
& \multicolumn{2}{c}{\quad\textbf{SQD} $(n)$\quad}
& \quad(S)\quad
& \textbf{Quantum}
& \quad\textbf{Classical}\quad \\

\midrule
\midrule

$24$ & $2^{24}$
& $\mathcal{O}(2^{9})$
& $326$ & $48)$
& $2^{12}$
& -45.488(312)
& -45.488 \\

$28$ & $2^{28}$
& $\mathcal{O}(2^{17})$
& $623$ & $72)$
& $2^{14}$
& -51.828(325)
& -51.828 \\

$32$ & $2^{32}$
& $\mathcal{O}(2^{22})$
& $938$ & $119)$
& $2^{16}$
& -56.543(269)
& -56.543 \\

$36$ & $2^{36}$
& $\mathcal{O}(2^{26})$
& $946$ & $153)$
& $2^{18}$
& -56.543(288)
& -56.543 \\

$40$ & $2^{40}$
& $\mathcal{O}(2^{29})$
& $2601$ & $232)$
& $2^{20}$
& -121.564(1610)
& -96.328\footnote{Accuracy is limited by the available memory.} \\

$44$ & $2^{44}$
& $\mathcal{O}(2^{33})$
& $3821$ & $389)$
& $2^{21}\footnote[2]{The shots could not be increased further due to the limited hardware resource.}$
& -103.261(8452)
& OOM \\

$48$ & $2^{48}$
& $\mathcal{O}(2^{35})$
& $3265$ & $982)$
& $2^{21}\footnotemark[2]$
& -102.445(7432)
& OOM \\

\bottomrule
\end{tabular}

\end{table*}

Figure~\ref{fig:quantumadvflowchart} illustrates the time-to-solution for $^{32}$Mg across these classical and quantum methods as the configuration space scales, clearly demonstrating the onset of the potential advantages within the resources considered. We compare the IBM Heron-class and Eagle-class quantum hardware, finding that the Heron-class hardware is nearly 3.5 times faster than the Eagle-class hardware~\cite{mayo2026benchmarkingquantumcomputersprotocols}. Given the current state of quantum hardware, the Heron family architectures can generate reliable samples from computationally hard circuits~\cite{martiel2026samplinghardcircuitsverifiably}. We implement SQD using the Heron Fez, via Qiskit~\cite{qiskit2024}. This SQD implementation shows a clear advantage over other methods, particularly when a large number of bases is used. The VQE method generally scales better than the classical eigensolvers eig and eigsh methods, converging quickly, but it fails to reach the standard values due to the implementation of deep circuits. For $44$ spin orbitals and above, VQE could outperform classical methods. Due to our RAM and QPU access limitations, we could not obtain results for more than 32 spin orbitals using VQE. 

The SQD approach not only depicts an advantage for cases with $40$ or more spin-orbitals, but also overcomes the RAM limitations of $768$~GB that restrict our Lanczos calculations. The SQD results are in excellent agreement with the classical calculations for model spaces containing up to 36 spin-orbitals. For the $40$ spin-orbitals case, the SQD method yields a ground-state energy closer to the experimental value than the classical calculation. This discrepancy with the classical result is due to the available system memory ($768$~GB RAM) that limits the number of vectors that can be retained during the classical diagonalization, thereby reducing the accuracy of the classical result. For larger model spaces ($44$ and $48$ spin-orbitals), exact classical diagonalization becomes infeasible due to memory limitations, whereas the quantum approach continues to provide approximate solutions. 

However, the finite number of measurement shots and the limited QPU execution time introduce statistical sampling errors, preventing the SQD algorithm from fully converging. The linear trend of SQD observed in Fig.~\ref{fig:quantumadvflowchart} arises from the increase in the number of measurement shots required for accurate sampling as the size of the model space grows. For small systems up to 24 spin-orbitals, the standard minimum of 1024 shots is sufficient for the SQD algorithm to converge reliably, and the runtime remains approximately constant. The larger shots taken to sample up to 24 to meet the basic standard of 1024 shots lead to the deviations (kinks) around 26 spin-orbitals from the otherwise linear trend. As the number of spin-orbitals increases, however, the Hilbert space expands exponentially, necessitating a corresponding increase in the number of measurement shots to adequately sample the quantum state. 
Consequently, the shot count was increased to approximately $2^{20}$ for the $40$ spin-orbitals calculation, which was sufficient for the SQD energy to converge close to the experimental value.  However, in calculations with 44 and 48 spin-orbitals, the SQD results did not fully converge due to the limited shot count.  The larger number of measurement shots and the QPU execution time required for full convergence are beyond our present resources, but are available in the market. 
 
Despite the above limitations, we demonstrated that the SQD avoids severe memory bottlenecks and longer execution times compared to other eigensolvers. SQD could be regarded as the current best among hybrid quantum algorithms and is highly efficient for large-basis cases, even surpassing the Lanczos method, which is widely used for nuclear energy-level calculations. SQD can also provide a few nearby energy levels, in addition to the ground state. The calculations presented here are resilient to QPU hardware noise and scalable to any HPC-QPU environment, establishing SQD as a leading algorithmic candidate for large-basis nuclear structure simulations and similar large-scale configuration interaction calculations in other areas of research. 

\textit{Conclusions}---The large-scale shell models to describe atomic nuclei have been quite successful except for the cases where the large configuration space involved imposes severe computational constraints.  $^{32}$Mg is one such case, with several interesting physics aspects that demand a rigorous calculation with a Hilbert space size up to $2^{48}$. While classical computers struggle to diagonalize a $2^{48}\times2^{48}$ matrix, it is possible to encode the information in $48$ qubits to evaluate $2^{48}$ amplitudes and utilize relevant quantum algorithms.  Most of such algorithms are impractical due to the inherent noise in the current noisy intermediate-scale quantum (NISQ) hardware.  We have demonstrated that we can overcome all these challenges by employing the sample-based quantum diagonalization method (SQD) and obtain the ground-state energy closer to the experimental value.  By using the quantum hardware solely as a configuration sampler and performing the actual exact diagonalization on a classical computer, SQD bypasses the noise accumulation typical of deep quantum circuits. Simultaneously, it overcomes the memory limits and time requirements of traditional classical solvers by truncating the basis space before diagonalization.  Our results, based on a given classical computing system coupled with a limited-resource cloud quantum computer, demonstrate the scalability of our SQD-based approach and provide a pathway to more rigorous calculations either with more computing resources or with resource-efficient protocols \cite{siwach2021, Singh_Nifeeya_shellmodel_GC}.  These hybrid quantum-centric methodologies might also be beneficial for a wider variety of problems involving larger matrices in other areas of research.

\textit{Acknowledgments}---We thank Pooja Siwach for fruitful discussions and comments on this manuscript.  This work is supported by the SERB-DST, Govt.~of India, via project \sloppy{CRG/2022/009359}.
We acknowledge the National Supercomputing Mission (NSM) for providing computing resources of `PARAM Ganga' at IIT Roorkee, which is implemented by C-DAC and supported by MeitY and DST, Govt.~of India.

\bibliography{Bibliography}

@article{troyer2005computational,
  title = {Computational Complexity and Fundamental Limitations to Fermionic Quantum Monte Carlo Simulations},
  author = {Troyer, Matthias and Wiese, Uwe Jens},
  journal = {Phys. Rev. Lett.},
  volume = {94},
  issue = {17},
  pages = {170201},
  numpages = {4},
  year = {2005},
  month = {May},
  publisher = {American Physical Society},
  doi = {10.1103/PhysRevLett.94.170201},
  url = {https://link.aps.org/doi/10.1103/PhysRevLett.94.170201}
}

@article{lanczos1950iteration,
    author = "Lanczos, Cornelius",
    title = "{An iteration method for the solution of the eigenvalue problem of linear differential and integral operators}",
    doi = "10.6028/jres.045.026",
    journal = "J. Res. Natl. Bur. Stand. B",
    volume = "45",
    pages = "255--282",
    year = "1950"
}

@article{Duriez2026,
  author  = {Duriez, Alan and Carvalho, Pamela C. and Barroca, Marco Antonio and Zipoli, Federico and Jaderberg, Ben and Neumann Barros Ferreira, Rodrigo and Sharma, Kunal and Mezzacapo, Antonio and Wunsch, Benjamin and Steiner, Mathias},
  title   = {Computing band gaps of periodic materials via sample-based quantum diagonalization},
  journal = {npj Comp. Mater.},
  year    = {2026},
  volume = {12},
  pages = {271},
  doi     = {10.1038/s41524-026-02059-0},
  issn    = {2057-3960}
}

@article{wietek2025xdiag,
  author = {Wietek, Alexander and Staszewski, Luke and Ulaga, Martin and Ebert, Paul L. and Karlsson, Hannes and Sarkar, Siddhartha and Shackleton, Leyna and Sinha, Aritra and Soares, Rafael D.},
  title = {XDiag: Exact diagonalization for quantum many-body systems},
  journal = {SciPost Phys. Codebases},
  volume = {70},
  year = {2026},
  doi = {10.21468/SciPostPhysCodeb.70},
  url = {https://scipost.org/SciPostPhysCodeb.70-r0.4},
}

@article{nandy2025quantum,
  author  = {Nandy, Pratik and Matsoukas-Roubeas, Apollonas S. and
             Mart{\'i}nez Azcona, Pablo and Dymarsky, Anatoly and
             {del Campo}, Adolfo},
  title   = {Quantum dynamics in Krylov space: Methods and applications},
  journal = {Phys. Rep.},
  volume  = {1125-1128},
  pages   = {1-82},
  year    = {2025},
  publisher = {Elsevier},
  doi     = {10.1016/j.physrep.2025.05.001}
}

@article{georgescu2014quantum,
  title = {Quantum simulation},
  author = {Georgescu, I. M. and Ashhab, S. and Nori, Franco},
  journal = {Rev. Mod. Phys.},
  volume = {86},
  issue = {1},
  pages = {153--185},
  numpages = {33},
  year = {2014},
  month = {Mar},
  publisher = {American Physical Society},
  doi = {10.1103/RevModPhys.86.153},
  url = {https://link.aps.org/doi/10.1103/RevModPhys.86.153}
}

@article{schollwock2005density,
  title = {The density-matrix renormalization group},
  author = {Schollw\"ock, U.},
  journal = {Rev. Mod. Phys.},
  volume = {77},
  issue = {1},
  pages = {259--315},
  numpages = {0},
  year = {2005},
  month = {Apr},
  publisher = {American Physical Society},
  doi = {10.1103/RevModPhys.77.259},
  url = {https://link.aps.org/doi/10.1103/RevModPhys.77.259}
}

@article{noack2005diagonalization,
    author = {Noack, Reinhard M. and Manmana, Salvatore R.},
    title = {Diagonalization‐ and Numerical Renormalization‐Group‐Based Methods for Interacting Quantum Systems},
    journal = {AIP Conf. Proc.},
    volume = {789},
    number = {1},
    pages = {93-163},
    year = {2005},
    month = {09},
    issn = {0094-243X},
    doi = {10.1063/1.2080349},
}

@article{fauseweh2024quantum,
author={Fauseweh, Benedikt},
title={Quantum many-body simulations on digital quantum computers: State-of-the-art and future challenges},
journal="Nat. Commun.",
year={2024},
month={Mar},
day={08},
volume={15},
number={1},
pages={2123},
issn={2041-1723},
doi={10.1038/s41467-024-46402-9},
url={https://doi.org/10.1038/s41467-024-46402-9}
}

@article{nyman2001iterative,
  title={Iterative diagonalization of a large sparse matrix using spectral transformation and filter diagonalization},
  author={Nyman, Gunnar and Yu, Hua Gen},
  journal={J. Comput. Methods Sci. Eng.},
  volume={1},
  number={2-3},
  pages={229--250},
  year={2001},
  publisher={SAGE Publications Sage UK: London, England},
  doi={https://doi.org/10.3233/JCM-2001-12-306
  }
}

@article{utsuno1999varying,
  title = {Varying shell gap and deformation in $\textit{N}\ensuremath{\sim}20$ unstable nuclei studied by the {Monte Carlo} shell model},
  author = {Utsuno, Yutaka and Otsuka, Takaharu and Mizusaki, Takahiro and Honma, Michio},
  journal = {Phys. Rev. C},
  volume = {60},
  issue = {5},
  pages = {054315},
  numpages = {8},
  year = {1999},
  month = {Oct},
  publisher = {American Physical Society},
  doi = {10.1103/PhysRevC.60.054315},
  url = {https://link.aps.org/doi/10.1103/PhysRevC.60.054315}
}

@article{shimizu2019thick,
  title={Thick-restart block Lanczos method for large-scale shell-model calculations},
  author={Shimizu, Noritaka and Mizusaki, Takahiro and Utsuno, Yutaka and Tsunoda, Yusuke},
  journal={Comput. Phys. Commun.},
  volume={244},
  pages={372--384},
  year={2019},
  publisher={Elsevier},
  doi={https://doi.org/10.1016/j.cpc.2019.06.011}
}

@article{kitamura2021coexisting,
  title={Coexisting normal and intruder configurations in $^{32}\text{Mg}$},
  author={Kitamura, N and Wimmer, K and Poves, Alfredo and Shimizu, N and Tostevin, JA and Bader, VM and Bancroft, C and Barofsky, D and Baugher, T and Bazin, D and et al.},
  journal={Phys. Lett. B},
  volume={822},
  pages={136682},
  year={2021},
  publisher={Elsevier},
  doi={https://doi.org/10.1016/j.physletb.2021.136682}
}

@misc{patra2026towards,
      title={Towards Chemically Accurate and Scalable Quantum Simulations on IQM Quantum Hardware: A Quantum-HPC Hybrid Approach}, 
      author={Anurag K. S. V. and Ashish Kumar Patra and Manas Mukherjee and Alok Shukla and Sai Shankar P. and Ruchika Bhat and Radhika T. S. L. and Jaiganesh G},
      year={2026},
      eprint={2604.01983},
      archivePrefix={arXiv},
      primaryClass={quant-ph},
      url={https://arxiv.org/abs/2604.01983}, 
}

@article{Warburton1990_32region_29to44,
  title = {Mass systematics for $\textit{A}=29-44$ nuclei: The deformed $\textit{A}\ensuremath{\sim}32$ region},
  author = {Warburton, E. K. and Becker, J. A. and Brown, B. A.},
  journal = {Phys. Rev. C},
  volume = {41},
  issue = {3},
  pages = {1147--1166},
  numpages = {0},
  year = {1990},
  month = {Mar},
  publisher = {American Physical Society},
  doi = {10.1103/PhysRevC.41.1147},
  url = {https://link.aps.org/doi/10.1103/PhysRevC.41.1147}
}

@article{caurier2005shellmodel,
  title = {The shell model as a unified view of nuclear structure},
  author = {Caurier, E. and Mart\'{\i}nez Pinedo, G. and Nowacki, F. and Poves, A. and Zuker, A. P.},
  journal = {Rev. Mod. Phys.},
  volume = {77},
  issue = {2},
  pages = {427--488},
  numpages = {0},
  year = {2005},
  month = {Jun},
  publisher = {American Physical Society},
  doi = {10.1103/RevModPhys.77.427},
  url = {https://link.aps.org/doi/10.1103/RevModPhys.77.427}
}

@article{Modine1996truncation,
  title = {Variational Hilbert-space-truncation approach to quantum Heisenberg antiferromagnets on frustrated clusters},
  author = {Modine, N. A. and Kaxiras, Efthimios},
  journal = {Phys. Rev. B},
  volume = {53},
  issue = {5},
  pages = {2546--2555},
  numpages = {0},
  year = {1996},
  month = {Feb},
  publisher = {American Physical Society},
  doi = {10.1103/PhysRevB.53.2546},
  url = {https://link.aps.org/doi/10.1103/PhysRevB.53.2546}
}

@article{hergert2020abinitio,
  author  = {Hergert, Heiko},
  title   = {A Guided Tour of {ab initio} Nuclear Many-Body Theory},
  journal = {Front. Phys.},
  volume  = {8},
  pages   = {379},
  year    = {2020},
  doi     = {10.3389/fphy.2020.00379}
}

@article{BARRETT2013131,
title = {ab initio no core shell model},
journal = {Prog. Part. Nucl. Phys.},
volume = {69},
pages = {131-181},
year = {2013},
issn = {0146-6410},
doi = {https://doi.org/10.1016/j.ppnp.2012.10.003},
url = {https://www.sciencedirect.com/science/article/pii/S0146641012001184},
author = {Bruce R. Barrett and Petr Navrátil and James P. Vary}
}

@article{otsuka2005tensorforce,
  title = {Evolution of Nuclear Shells due to the Tensor Force},
  author = {Otsuka, Takaharu and Suzuki, Toshio and Fujimoto, Rintaro and Grawe, Hubert and Akaishi, Yoshinori},
  journal = {Phys. Rev. Lett.},
  volume = {95},
  issue = {23},
  pages = {232502},
  numpages = {4},
  year = {2005},
  month = {Nov},
  publisher = {American Physical Society},
  doi = {10.1103/PhysRevLett.95.232502},
  url = {https://link.aps.org/doi/10.1103/PhysRevLett.95.232502}
}

@article{fetter_2003_Quantum_theory_of_many_particle_systems,
author = {Fetter, Alexander and Walecka, J. and Kadanoff, Leo},
year = {2008},
month = {12},
pages = {54-55},
title = {Quantum Theory of Many Particle Systems},
volume = {25},
journal = {Phys. Today},
doi = {10.1063/1.3071096}
}

@article{Whitfield10032011,
    author = {James D. Whitfield and Jacob Biamonte and Alán Aspuru Guzik},
    title = {Simulation of electronic structure Hamiltonians using quantum computers},
    journal = {Mol. Phys.},
    volume = {109},
    number = {5},
    pages = {735--750},
    year = {2011},
    publisher = {Taylor \& Francis},
    doi = {10.1080/00268976.2011.552441}
}

@article{Bartlett_Rodney2006_New_perspectives_on_unitary_coupled_cluster_theory,
author = {Taube, Andrew G. and Bartlett, Rodney J.},
title = {New perspectives on unitary coupled-cluster theory},
journal = {Int. J. of Quantum Chem.},
volume = {106},
number = {15},
pages = {3393-3401},
doi = {https://doi.org/10.1002/qua.21198},
year = {2006}
}

@article{siwach2021,
  title = {Quantum simulation of nuclear Hamiltonian with a generalized transformation for Gray code encoding},
  author = {Siwach, Pooja and Arumugam, P.},
  journal = {Phys. Rev. C},
  volume = {104},
  issue = {3},
  pages = {034301},
  numpages = {13},
  year = {2021},
  month = {Sep},
  publisher = {American Physical Society},
  doi = {10.1103/PhysRevC.104.034301},
  url = {https://link.aps.org/doi/10.1103/PhysRevC.104.034301}
}

@misc{pandey2026rvvqe,
      author={Durgesh Pandey and Ankit Kumar Das and P. Arumugam},
      title={{Real Variance-Based Variational Quantum Eigensolver for Non-Hermitian Matrices}}, 
      year={2026},
      eprint={2603.28892},
      archivePrefix={arXiv},
      primaryClass={quant-ph},
      url={https://arxiv.org/abs/2603.28892}, 
}

@article{Dumitrescu2018Cloud_Quantum_Computing_of_an_Atomic_Nucleus,
  title = {Cloud Quantum Computing of an Atomic Nucleus},
  author = {Dumitrescu, E. F. and McCaskey, A. J. and Hagen, G. and Jansen, G. R. and Morris, T. D. and Papenbrock, T. and Pooser, R. C. and Dean, D. J. and Lougovski, P.},
  journal = {Phys. Rev. Lett.},
  volume = {120},
  issue = {21},
  pages = {210501},
  numpages = {6},
  year = {2018},
  month = {May},
  publisher = {American Physical Society},
  doi = {10.1103/PhysRevLett.120.210501},
  url = {https://link.aps.org/doi/10.1103/PhysRevLett.120.210501}
}

@article{JordanWigner1928,
  author    = {Jordan, P. and Wigner, E.},
  title     = {{\"U}ber das Paulische {\"A}quivalenzverbot},
  journal   = {Z. Phys.},
  year      = {1928},
  volume    = {47},
  number    = {9},
  pages     = {631--651},
  doi       = {10.1007/BF01331938},
  url       = {https://doi.org/10.1007/BF01331938}
}

@article{Batista2001,
  title = {Generalized {Jordan-Wigner} Transformations},
  author = {Batista, C. D. and Ortiz, G.},
  journal = {Phys. Rev. Lett.},
  volume = {86},
  issue = {6},
  pages = {1082--1085},
  numpages = {0},
  year = {2001},
  month = {Feb},
  publisher = {American Physical Society},
  doi = {10.1103/PhysRevLett.86.1082},
  url = {https://link.aps.org/doi/10.1103/PhysRevLett.86.1082}
}

@article{peruzzo2014variational,
  title={A variational eigenvalue solver on a photonic quantum processor},
  author={Peruzzo, A. and McClean, J. R. and Shadbolt, P. and Yung, M.H. and Zhou, X.Q. and Love, P. J and Aspuru Guzik, A. and O'Brien, J. L.},
  journal="Nat. Commun.",
  volume={5},
  pages={4213},
  year={2014},
  publisher={Nature Publishing Group},
  doi = {10.1038/ncomms5213}
}

@article{Barison_2025_SQD,
    doi = {10.1088/2058-9565/adb781},
    url = {https://doi.org/10.1088/2058-9565/adb781},
    year = {2025},
    month = {feb},
    publisher = {IOP Publishing},
    volume = {10},
    number = {2},
    pages = {025034},
    author = {Barison, Stefano and Robledo Moreno, Javier and Motta, Mario},
    title = {Quantum-centric computation of molecular excited states with extended sample-based quantum diagonalization},
    journal = {Quantum Sci. and Technol.}
}

@article{RichterUSDA_USDB,
  title = {$\mathit{sd}$-shell observables for the {USDA} and {USDB} Hamiltonians},
  author = {Richter, W. A. and Mkhize, S. and Brown, B. Alex},
  journal = {Phys. Rev. C},
  volume = {78},
  issue = {6},
  pages = {064302},
  numpages = {7},
  year = {2008},
  month = {Dec},
  publisher = {American Physical Society},
  doi = {10.1103/PhysRevC.78.064302},
  url = {https://link.aps.org/doi/10.1103/PhysRevC.78.064302}
}

@misc{martiel2026samplinghardcircuitsverifiably,
      title={Sampling hard circuits with verifiably high fidelity}, 
      author={Simon Martiel and Jay U Chung and Alireza Seif and Soumik Ghosh and Ian Hincks and Abhinav Deshpande and Bill Fefferman and Jay M. Gambetta and Ali Javadi Abhari},
      year={2026},
      eprint={2607.25941},
      archivePrefix={arXiv},
      primaryClass={quant-ph},
      url={https://arxiv.org/abs/2607.25941}, 
}

@article{Romero_2019_UCCansatz,
    doi = {10.1088/2058-9565/aad3e4},
    url = {https://doi.org/10.1088/2058-9565/aad3e4},
    year = {2018},
    month = {oct},
    publisher = {IOP Publishing},
    volume = {4},
    number = {1},
    pages = {014008},
    author = {Romero, Jonathan and Babbush, Ryan and McClean, Jarrod R and Hempel, Cornelius and Love, Peter J and Aspuru Guzik, Alán},
    title = {Strategies for quantum computing molecular energies using the unitary coupled cluster ansatz},
    journal = {Quantum Sci. Technol.}
}

@article{Perez-Obiol2023Nuclear_shell-model_simulation_in_digital_quantum_computers,
    author={P{\'e}rez Obiol, A.
    and Romero, A. M.
    and Men{\'e}ndez, J.
    and Rios, A.
    and Garc{\'i}a S{\'a}ez, A.
    and Juli{\'a} D{\'i}az, B.},
    title={Nuclear shell-model simulation in digital quantum computers},
    journal={Sci. Rep.},
    year={2023},
    month={Jul},
    day={29},
    volume={13},
    number={1},
    pages={12291},
    issn={2045-2322},
    doi={10.1038/s41598-023-39263-7},
    url={https://doi.org/10.1038/s41598-023-39263-7}
    }

@article{SQD_2025_Chemistry_beyond_the_scale_of_exact_diagonalization_on_a_quantum-centric_supercomputer,
author = {Javier Robledo Moreno  and Mario Motta  and Holger Haas  and Ali Javadi-Abhari  and Petar Jurcevic  and William Kirby  and Simon Martiel  and Kunal Sharma  and Sandeep Sharma  and Tomonori Shirakawa  and Iskandar Sitdikov  and Rong Yang Sun  and Kevin J. Sung  and Maika Takita  and Minh C. Tran  and Seiji Yunoki  and Antonio Mezzacapo },
title = {Chemistry beyond the scale of exact diagonalization on a quantum-centric supercomputer},
journal = {Sci. Adv.},
volume = {11},
number = {25},
pages = {eadu9991},
year = {2025},
doi = {10.1126/sciadv.adu9991},
URL = {https://www.science.org/doi/abs/10.1126/sciadv.adu9991}
}

@article{Higgott2019variationalquantumdeflation,
  doi = {10.22331/q-2019-07-01-156},
  url = {https://doi.org/10.22331/q-2019-07-01-156},
  title = {Variational {Q}uantum {C}omputation of {E}xcited {S}tates},
  author = {Higgott, Oscar and Wang, Daochen and Brierley, Stephen},
  journal = {{Quantum}},
  issn = {2521-327X},
  publisher = {{Verein zur F{\"{o}}rderung des Open Access Publizierens in den Quantenwissenschaften}},
  volume = {3},
  pages = {156},
  month = jul,
  year = {2019}
}

@article{Huang_2021,
doi = {10.1088/1674-1137/abddb0},
url = {https://doi.org/10.1088/1674-1137/abddb0},
year = {2021},
month = {mar},
publisher = {Chinese Physical Society and the Institute of High Energy Physics of the Chinese Academy of Sciences and the Institute of Modern Physics of the Chinese Academy of Sciences and IOP Publishing Ltd},
volume = {45},
number = {3},
pages = {030002},
author = {Huang, W.J. and Wang, Meng and Kondev, F.G. and Audi, G. and Naimi, S.},
title = {The \text{AME} 2020 atomic mass evaluation (I). Evaluation of input data, and adjustment procedures},
journal = {Chin. Phys. C}
}

@article{Wang_2021_II,
doi = {10.1088/1674-1137/abddaf},
url = {https://doi.org/10.1088/1674-1137/abddaf},
year = {2021},
month = {mar},
publisher = {Chinese Physical Society and the Institute of High Energy Physics of the Chinese Academy of Sciences and the Institute of Modern Physics of the Chinese Academy of Sciences and IOP Publishing Ltd},
volume = {45},
number = {3},
pages = {030003},
author = {Wang, Meng and Huang, W.J. and Kondev, F.G. and Audi, G. and Naimi, S.},
title = {The \text{AME} 2020 atomic mass evaluation (II). Tables, graphs and references},
journal = {Chin. Phys. C}
}

@article{Nakanishi2019_SSVQE,
author = {Nakanishi, Ken and Mitarai, Kosuke and Fujii, Keisuke},
year = {2019},
month = {10},
pages = {},
title = {Subspace-search variational quantum eigensolver for excited states},
volume = {1},
journal = {Phys. Rev. Res.},
doi = {10.1103/PhysRevResearch.1.033062}
}

@article{Danbo20222VVQE,
author = {Danbo, Zhang and Chen, Bin Lin and Yuan, Zhan Hao and Yin, Tao},
year = {2022},
month = {11},
pages = {},
title = {Variational quantum eigensolvers by variance minimization},
volume = {31},
journal = {Chinese Phys. B},
doi = {10.1088/1674-1056/ac8a8d}
}

@article{Huggins_2020_IOP_nonorthogonal_VQE_solver,
    doi = {10.1088/1367-2630/ab867b},
    url = {https://doi.org/10.1088/1367-2630/ab867b},
    year = {2020},
    month = {jul},
    publisher = {IOP Publishing},
    volume = {22},
    number = {7},
    pages = {073009},
    author = {Huggins, William J and Lee, Joonho and Baek, Unpil and O’Gorman, Bryan and Whaley, K Birgitta},
    title = {A non-orthogonal variational quantum eigensolver},
    journal = {New J. Phys.}
}

@article{Wang2021_barren_plateaus,
author={Wang, Samson
and Fontana, Enrico
and Cerezo, M.
and Sharma, Kunal
and Sone, Akira
and Cincio, Lukasz
and Coles, Patrick J.},
title={Noise-induced barren plateaus in variational quantum algorithms},
journal="Nat. Commun.",
year={2021},
month={Nov},
day={29},
volume={12},
number={1},
pages={6961},
issn={2041-1723},
doi={10.1038/s41467-021-27045-6},
url={https://doi.org/10.1038/s41467-021-27045-6}
}

@article{Singh_Nifeeya_shellmodel_GC,
  title = {Advancing quantum simulations of the nuclear shell model with Gray-code--based resource-efficient protocols},
  author = {Singh, Nifeeya and Siwach, Pooja and Arumugam, P.},
  journal = {Phys. Rev. C},
  volume = {112},
  issue = {3},
  pages = {034320},
  numpages = {20},
  year = {2025},
  month = {Sep},
  publisher = {American Physical Society},
  doi = {10.1103/bbkf-fjxj},
  url = {https://link.aps.org/doi/10.1103/bbkf-fjxj}
}

@misc{Lanczos__method,
      author={Noah Amsel and Tyler Chen and Anne Greenbaum and Cameron Musco and Chris Musco},
      title={Nearly Optimal Approximation of Matrix Functions by the {L}anczos Method}, 
      year={2024},
      eprint={2303.03358},
      archivePrefix={arXiv},
      primaryClass={math.NA},
      url={https://arxiv.org/abs/2303.03358}, 
}

@ARTICLE{2020SciPy,
  author  = {Virtanen, Pauli and Gommers, Ralf and Oliphant, Travis E. and
            Haberland, Matt and Reddy, Tyler and Cournapeau, David and
            Burovski, Evgeni and Peterson, Pearu and Weckesser, Warren and
            Bright, Jonathan and {van der Walt}, St{\'e}fan J. and
            Brett, Matthew and Wilson, Joshua and Millman, K. Jarrod and
            Mayorov, Nikolay and Nelson, Andrew R. J. and Jones, Eric and
            Kern, Robert and Larson, Eric and Carey, C J and
            Polat, {\.I}lhan and Feng, Yu and Moore, Eric W. and
            {VanderPlas}, Jake and Laxalde, Denis and Perktold, Josef and
            Cimrman, Robert and Henriksen, Ian and Quintero, E. A. and
            Harris, Charles R. and Archibald, Anne M. and
            Ribeiro, Ant{\^o}nio H. and Pedregosa, Fabian and
            {van Mulbregt}, Paul and {SciPy 1.0 Contributors}},
  title   = {{{SciPy} 1.0: Fundamental Algorithms for Scientific
            Computing in Python}},
  journal = {Nat. Methods},
  year    = {2020},
  volume  = {17},
  pages   = {261--272},
  adsurl  = {https://rdcu.be/b08Wh},
  doi     = {10.1038/s41592-019-0686-2},
}

@misc{benstead2026QPE32Mg,
      title={Fault-tolerant quantum algorithms for simulating atomic nuclei}, 
      author={James Benstead and Michael Garn and Neil Gaspar and Sean Greenaway and Angus Kan and Lloyd La Ronde and Chandan Sarma and Paul Stevenson},
      year={2026},
      eprint={2607.21563},
      archivePrefix={arXiv},
      primaryClass={quant-ph},
      url={https://arxiv.org/abs/2607.21563}, 
}

@misc{PoojaRao_model_timecalc2026,
      title={Performance Model for Hybrid Quantum-Classical Workflows}, 
      author={Pooja Rao and Dimitar Trenev and Jerome Gonthier and Taylor Patti and Sebastian Stern and Tyler Takeshita and Yuri Alexeev and Cedric Lin and Sam McArdle and Justin Lietz and Katherine Klymko and Ermal Rrapaj and Norm Tubman and Krysta Svore and Peter Komar and Elica Kyoseva},
      year={2026},
      eprint={2607.15426},
      archivePrefix={arXiv},
      primaryClass={quant-ph},
      url={https://arxiv.org/abs/2607.15426}, 
}

@misc{qiskit2024,
      title={Quantum computing with {Q}iskit},
      author={Javadi Abhari, Ali and Treinish, Matthew and Krsulich, Kevin and Wood, Christopher J. and Lishman, Jake and Gacon, Julien and Martiel, Simon and Nation, Paul D. and Bishop, Lev S. and Cross, Andrew W. and Johnson, Blake R. and Gambetta, Jay M.},
      year={2024},
      doi={10.48550/arXiv.2405.08810},
      eprint={2405.08810},
      archivePrefix={arXiv},
      primaryClass={quant-ph}
}

@misc{mayo2026benchmarkingquantumcomputersprotocols,
      title={Benchmarking Quantum Computers via Protocols, Comparing {IBM}'s Heron vs {IBM}'s Eagle}, 
      author={Nitay Mayo and Tal Mor and Yossi Weinstein},
      year={2026},
      eprint={2603.04377},
      archivePrefix={arXiv},
      primaryClass={quant-ph},
      url={https://arxiv.org/abs/2603.04377}, 
}

@misc{yoshida2026nuclearmanybodysystemsQPEQKrylov,
      title={Nuclear Many-Body Systems as Benchmarks for Quantum Computing}, 
      author={Sota Yoshida and Alessandro Baroni and Takayuki Miyagi and Ermal Rrapaj},
      year={2026},
      eprint={2607.08047},
      archivePrefix={arXiv},
      primaryClass={quant-ph},
      url={https://arxiv.org/abs/2607.08047}, 
}

@article{AJAGEKAR2020106630,
  title = {Quantum computing based hybrid solution strategies for large-scale discrete-continuous optimization problems},
  author = {Akshay Ajagekar and Travis Humble and Fengqi You},
  journal = {Comput. \& Chem. Eng.},
  volume = {132},
  pages = {106630},
  year = {2020},
  doi = {10.1016/j.compchemeng.2019.106630},
  url = {https://www.sciencedirect.com/science/article/pii/S0098135419307665#cited-by}
}

@misc{hep_quantum_advantage,
  author = {Il{\v{c}}i{\'c}, Fran and Majumdar, Ritajit and Mathew, Emil and Earnest Noble, Nathan and Raychowdhury, Indrakshi},
  title = {Observation of Robust and Coherent Non-Abelian Hadron Dynamics on Noisy Quantum Processors},
  year = {2026},
  eprint = {2602.18080},
  archivePrefix = {arXiv},
  primaryClass = {hep-lat},
  doi = {10.48550/arXiv.2602.18080},
  url = {https://arxiv.org/abs/2602.18080}
}

@misc{paramganga,
  author       = {{Institute Computer Centre, IIT Roorkee}},
  title        = {{PARAM Ganga: High-Performance Computing Facility}},
  year         = {2026},
  url          = {https://hpc.iitr.ac.in/}
}

@misc{pandey2026quGCM,
      title={Quantum Simulation of Nuclear Shell Model Using {GCM}-Based Methods on {NISQ} Devices}, 
      author={Durgesh Pandey and Ashutosh Singh and Ankit Kumar Das and P. Arumugam},
      year={2026},
      eprint={2608.01769},
      archivePrefix={arXiv},
      primaryClass={nucl-th},
      url={https://arxiv.org/abs/2608.01769}, 
}

@article{Cao_2025,
title = {{ab initio} study in the island of inversion within the two-major-shell valence space},
journal = {Phys. Lett. B},
volume = {871},
pages = {140034},
year = {2025},
issn = {0370-2693},
doi = {https://doi.org/10.1016/j.physletb.2025.140034},
url = {https://www.sciencedirect.com/science/article/pii/S0370269325007920},
author = {X.C. Cao and C.F. Jiao}
}

\end{document}